\documentclass[5p]{elsarticle}
\usepackage{graphicx}
\usepackage{graphics}
\usepackage{amssymb}
\usepackage{nicefrac}

\newcommand{\iso} [2] {\ensuremath{{}^{#2}\!{\mathrm{#1}}}}
\newcommand{\multiso} [3] {\ensuremath{{}^{#2}\!{\mathrm{#1}_{#3}}}}
\newcommand{\Cone}{\ensuremath{\mathrm{C1}}}
\newcommand{\Ctwo}{\ensuremath{\mathrm{C2}}}

\newcommand{\Tone}{\ensuremath{T\mathrm{_1}}}
\newcommand{\Ttwo}{\ensuremath{T\mathrm{_2}}}

\newcommand{\Hany}{\ensuremath{\mathrm{H}}}
\newcommand{\Htwo}{\ensuremath{\mathrm{H_{\alpha}}}}
\newcommand{\Hthree}{\ensuremath{\mathrm{H3}}}
\newcommand{\Hfour}{\ensuremath{\mathrm{H4}}}

\newcommand{\Alp}{\ensuremath{\mathrm{\alpha}}}
\newcommand{\DW}{\ensuremath{\mathrm{D_2O}}}
\newcommand{\Mag}{\ensuremath{\mathrm{Mag\-ne\-vist}}}
\newcommand{\Magr}{\Mag$^{\mathrm{\scriptsize\textregistered}}$}
\def        \sec          {\ensuremath{\mathrm{s}}}
\def        \msec         {\ensuremath{\mathrm{ms}}}
\def        \mM           {\ensuremath{\mathrm{mM}}}
\def        \Hz           {\ensuremath{\mathrm{Hz}}}

\def	    \eps  	  {\varepsilon}

\def	    \RR   {\ensuremath{\mathcal{R}}}

\begin{document}

\begin{frontmatter}

\title{Heat-Bath Cooling of Spins in Two Amino Acids}

\author[TecChem]{Y.~Elias\corref{cor1}}
\ead{ye1@technion.ac.il}

\author[TecChem]{H.~Gilboa}

\author[TecCS]{T.~Mor}

\author[TecPhys]{Y.~Weinstein}

\address[TecChem]{Schulich Faculty of Chemistry, Technion 32000, Haifa, Israel}
\address[TecCS]{Department of Computer Science, Technion 32000, Haifa, Israel}
\address[TecPhys]{Department of Physics, Technion 32000, Haifa, Israel}

\cortext[cor1]{Presented at the International Society for Magnetic Resonance
 (ISMAR) in Florence, 2010.}
\cortext[cor2]{Corresponding author. Fax: +972 4 8293900.}

\begin{abstract}
Heat-bath cooling is a component of practicable algorithmic cooling of spins,
 an approach which might be useful for in vivo \iso{C}{13} spectroscopy, in
 particular for prolonged metabolic processes where substrates that are 
hyperpolarized ex-vivo are not effective.
We applied heat-bath cooling to 1,2-\multiso{C}{13}{2}-amino acids, using the
 \Alp\ protons to shift entropy from selected carbons to the environment. For
 glutamate and glycine, both carbons were cooled by about $2.5$-fold,
and in other experiments the polarization of \Cone\ nearly doubled while all other
spins had equilibrium polarization, indicating reduction in total entropy.
 The effect of adding \Magr, a gadolinium contrast agent, on heat-bath cooling
 of glutamate was investigated.
\end{abstract}

\end{frontmatter}

\newpage

\section{Introduction}
%
%
\emph{Heat-bath cooling}~\cite{POTENT} is a key building-block of
 \emph{algorithmic cooling (AC)}, a recent spin-cooling approach originally
 proposed for purifying qubits in NMR quantum computers~\cite{BMRVV02, FLMR04}.
AC intersperses heat-bath cooling (which combines polarization transfer
 and fast repolarization) with reversible polarization
 compression. By itself, polarization compression~\cite{Sorensen89, SV99} is 
 limited due to the entropy bound (also called \emph{Shannon's
 bound})~\cite{CT91}, and further limited by the unitary bound on spin
 dynamics~\cite{Sorensen89}. However, heat-bath cooling allows AC to cool spins
 (spin-half nuclei) far beyond these bounds~\cite{BMRVV02, FLMR04, SMW05, EFMW06,
 SMW07, EMW11}. 
 AC, as well as its components, heat-bath cooling and polarization compression,
may enhance in vivo \iso{C}{13}
magnetic resonance spectroscopy (13C-MRS), which enables dynamic analysis of
metabolic processes in living orga\-ni\-sms~\cite{RLH+03, RC05}. 

%
Heat-bath cooling achieves controlled entropy extraction from suitable 
spin systems~\cite{POTENT} by applying one or more steps of \emph{selective-reset}.
 Each selective-reset cools a selected slow-relaxing spin
 (\emph{computation spin}) by polarization transfer (PT) from a more polarized
 \emph{reset spin}, and this PT is followed by a delay. We call this idle step
 (the delay) a \emph{reset step}. The reset spin repolarizes rapidly during that
 reset step so that the spin system is cooled; the reset spin may be reused
 via an iterative (or recursive) algorithm~\cite{BMRVV02, FLMR04}.
For a pair of computation and reset spins, respectively $c$ and $r$, we denote
 here by $\RR(c,r)$ the ratio between the thermalization time of the computation
 spin, \Tone(c), and the characteristic repolarization time of the reset spin
 (for liquid NMR, it is the thermalization time, \Tone(r)). A 
 large ratio, $\RR(c,r) \gg 1,$ is required to preserve most of the enhanced
 polarization of the computation spin during reset steps. In addition, strong
scalar J-coupling (or for solids dipolar coupling) is required, $J \gg 1/\Ttwo,$
such that loss of polarization by decoherence during PT would be minimized.  

%
AC produces spin-cooling using heat-bath cooling steps, wherein the spin
 entropy is transiently reduced (the
 spins are \emph{cooled}) even though no physical cooling is applied to the
 system. Practicable cooling algorithms that yield moderate cooling already for
 small spin systems were suggested~\cite{FLMR04, EFMW06}, and improved algorithms
 were recently derived~\cite{EMW11}. 
 Heat-bath cooling, \emph{by itself}, can already cool the entire spin system,
bypassing the entropy bound, if, at equilibrium, the reset spin polarization,
 $\eps(r),$ is higher than the polarization of the computation spins,
 $\eps(c)$~\cite{FLMR04}.
 
%
For larger spin systems, AC can theoretically generate hyperpolarized spins, for
 which the
polarization is increased by several orders of magnitude~\cite{BMRVV02, FLMR04,
 SMW05, EFMW06, SMW07, EMW11}. However, such extensive cooling faces
 significant obstacles and requires many selective-resets and is not expected
 in the near future.
Other spin-cooling methods that produce hyperpolarization include 
dynamic nuclear polarization in the solid state followed by rapid thawing
 (dissolution DNP)~\cite{AFGH+03}, parahydrogen
 induced polarization (PH\-IP)~\cite{BHL+05}, and optical pumping of
 noble gases~\cite{OS04}. For small spin systems, AC is applicable with
 current technology, and has two important advantages over
dissolution DNP, the most generally applicable hyperpolarization method: 
both AC and heat-bath cooling
 require no special apparatus and may therefore be applied
 directly in vivo, in particular to enhance 13C-MRS of slow metabolism in the
 brain; Furthermore, the polarization enhancement may be
 \emph{ap\-plied/re\-plen\-ished in vivo} after several hours.

%
Significant entropy reduction was already demonstrated
 experimentally~\cite{POTENT}. That heat-bath cooling experiment used 
 \multiso{C}{13}{2}-tri\-chlo\-ro\-ethylene (TCE) in chloroform solution (with
 $\eps(r) \sim 4\eps(c)$),
 a simple model system comprising three spins - two carbons and one
 proton, where \Cone\ is the quaternary carbon and \Ctwo\ is bound to the 
 proton.
 A paramagnetic reagent, Cr(acac)$_3$,
 improved both \Tone\
 ratios~\cite{FMW03}, such that $\RR(\Cone,\Hany) \sim 10$ and $\RR(\Ctwo,
 \Hany) \sim 5.$ 
The entire spin system was cooled beyond the entropy bound by a dual
 selective-reset, using the four-step
pulse sequence POTENT (POlarization Transfer by ENvironment Thermalization)
~\cite{POTENT}: 
\begin{enumerate}
\item PT \#1. $\Hany \rightarrow \iso{C}{13} \rightarrow \iso{C}{13}$
(\emph{HCC}) relay implemented by two refocused INEPT
 sequences~\cite{MF79, BE80} using non-selective (``hard'') pulses.
\item Reset \#1. A significant delay of $\sim2 \Tone(\Hany)$.
\item PT \#2. Refocused INEPT from the partly repolarized proton to \Ctwo.
\item Reset \#2. A significant delay of $\sim3 \Tone(\Hany)$.
\end{enumerate}
After POTENT, both \iso{C}{13} were cooled nearly
 two-fold, while the proton regained most of its equilibrium polarization,
such that the \emph{entire} spin system was cooled.
In the submitted version of~\cite{POTENT}, another heat-bath cooling experiment is
 described where 
only steps 1--3 above are performed, such that both carbons are cooled, 
each by a factor greater than two. 

%
 We chose to work here with glycine and glutamate, which are both
 neurotransmitters~\cite{NFP+03}; Glutamate has the major excitatory
 role in the human brain,
while glycine assumes an inhibitory function in specialized
 neurons~\cite{NFP+03} and also acts as a cofactor for excitatory
 neurotransmission within the forebrain~\cite{PFW+06}.
Glutamate is present in the brain at significant concentrations (as
 high as $10\ \mM$)~\cite{SRH+08}, and the progressive labeling of its
carbons during metabolism of \iso{C}{13}-enriched substrates (commonly
 glucose) was studied extensively~\cite{NFP+03, SRH+08,
 GAC+03, RBHS03, LSER03}. Reduced glutamate
 neurotransmission in the brain is associated with Alzheimer's
 disease~\cite{LSER03} and with neuropsychiatric disorders~\cite{SRH+08}.
Gly\-cine is normally present at much lower concentrations~\cite{NFP+03}, and
its measurement in vivo is therefore challenging. By proton MRS, the single
 proton signal of glycine was recently resolved in vivo from the overlapping
 signal of the much more abundant myo-inositol~\cite{PFW+06, CDH+09}.
In vivo 13C-MRS could also measure glycine, which is obtained in neurons from
 glucose (via serine)~\cite{NFP+03}.
 Abnormal levels of glycine are implicated in schizophrenia and other
 neuropsychiatric disorders~\cite{PFW+06}, and recently glycine was also
shown to be a biomarker of tumor malignancy~\cite{CDH+09}. 

\begin{figure}[h]
\begin{center}
\includegraphics[scale=0.66]{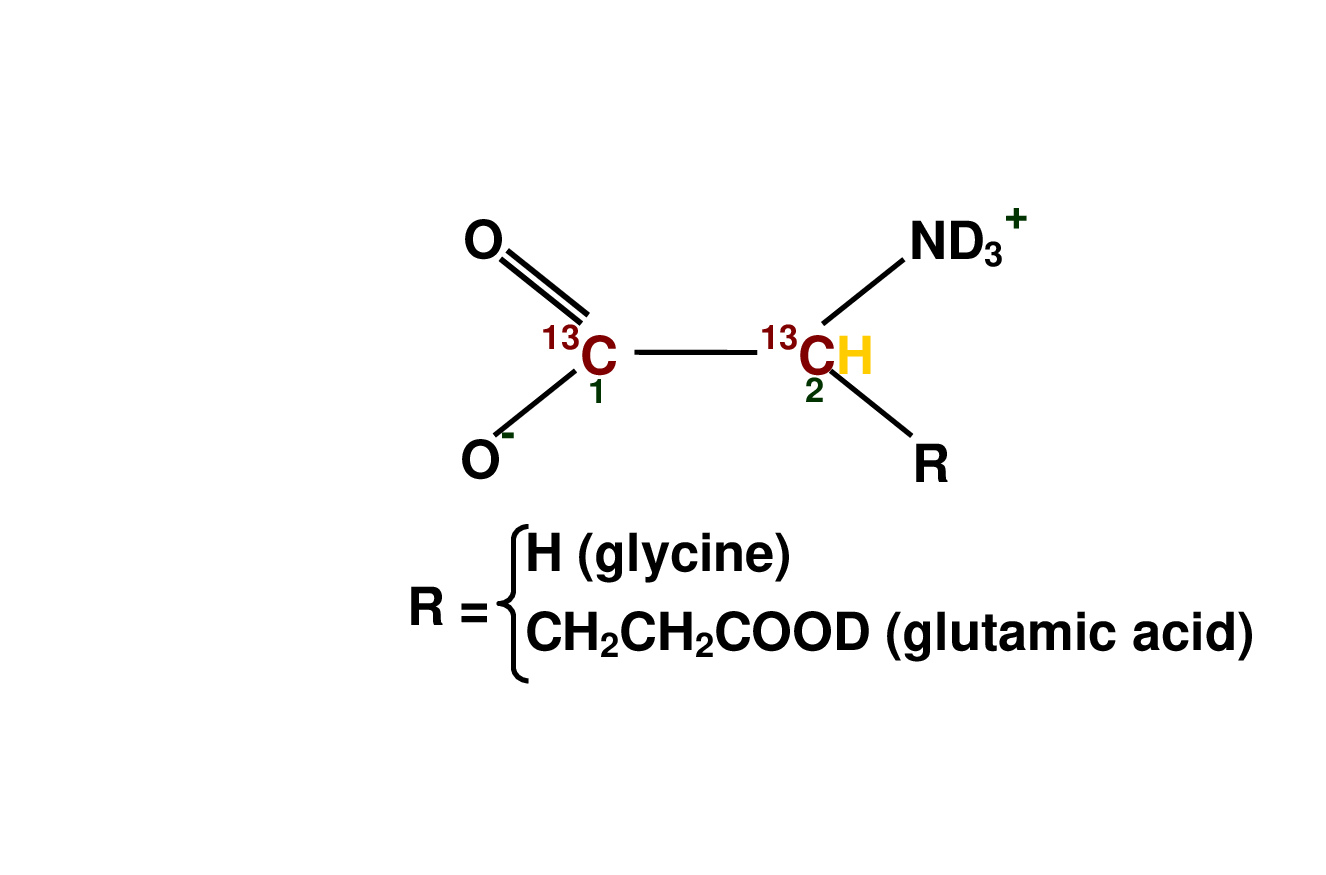}
\end{center}
\caption{
Amino acids used in heat-bath cooling experiments. 
}
\label{fig:amino}
\end{figure}

In this letter, we describe heat-bath cooling of the spin systems of glutamic
acid ($7$ spins: one \Alp\ proton, \iso{C}{13}1, \iso{C}{13}2, and 4 methylene
 protons) and glycine ($4$ spins: two \Alp\ protons, \iso{C}{13}1, and
 \iso{C}{13}2), depicted in Fig.~\ref{fig:amino}. For these systems, perfect PT
 from the reset proton(s) can increase 13C polarization by about
 4-fold~\cite{Sorensen89, BE80}, and sufficiently high $\RR(\Cone, \Htwo)$ were
 expected and confirmed. For each spin system, the
 entropy was reduced by a single selective-reset (the first two steps of
 POTENT), and in a different experiment \Cone\ and \Ctwo\ were cooled to a
 significant extent by adding a PT from the partially repolarized \Alp\
 proton(s) to \Ctwo; step 3 of POTENT.
Finally, we examined the possibility of improving the much lower \Tone\
ratio, $\RR(\Ctwo, \Htwo),$ in order to enable AC, by adding \Magr, a
common Gd-based MRI contrast agent~\cite{LEM06}.

\section{Materials and Methods}
%
%
NMR samples (0.5 mL) of \iso{C}{13}-labeled amino acids (99\% \iso{C}{13},
 Cam\-bridge Isotope Laboratories) were prepared in \DW\ (99.96\% D), and the pH
 was adjusted by adding K$_3$PO$_4$ (anhydrous, 99\% pure, Sigma-Ald\-rich) and
 measured using 6.4--8.0 pH sticks (0.2 units, Lyphan); see samples $G$ and $E$
in Table~\ref{tab:t1}.
\DW\ solutions of \Magr\ were prepared from lyo\-phi\-lized samples of a
 clinical aqueous solution (0.5M, Bayer HealthCare Pharmaceuticals). A dilute
 solution of \Mag\ in \DW\ ($5\ \mM$) was added directly to the NMR tubes with a
 micropipette in $5\ \mu$L aliquots, corresponding to incremental \Mag\
 concentrations of $0.05\ \mM$, similar to a recent study of glycine
 relaxation~\cite{GLE09}. Aliquots were added until $\Tone(\Htwo)$ was reduced
 to about $1 \sec$; see samples $E_M$ and $E_M'$ in Table~\ref{tab:t1}.

%
All experiments were performed on a Bruker Avance 500 spectrometer with
a standard 5 mm BBO probe and a digital variable temperature control unit,
at controlled temperatures ($0.1$K accuracy). \Tone\ relaxation times were
 measured by 
 inversion recovery (with the standard Bruker t1ir pulse sequence), consisting
 of a prolonged
 recycle delay ($7\Tone$), followed by $180^{\circ}, \tau, 90^{\circ},$ and
 acquisition. 17 loga\-rith\-mi\-cally-spaced $\tau$ values were used, such that a few were below \Tone\ and at least one delay exceeded $7\Tone$.

%
The POTENT pulse sequence~\cite{POTENT} and its components described above were 
used with two main modifications; First, the INEPT pulse sequences in steps 1
 and 3 were implemented with spin-selective low-power pulses for \iso{C}{13}
 excitation and inversion, which had shapes of E-BURP1 and I-BURP1~\cite{GF91},
 respectively, and a duration of $t=1~\msec.$ Second, during the PT from \Ctwo\
 to \Cone\ in step 1, WALTZ-16~\cite{SRFF83} proton decoupling was employed. 

\begin{figure}[h]
\begin{center}
\includegraphics[scale=0.66]{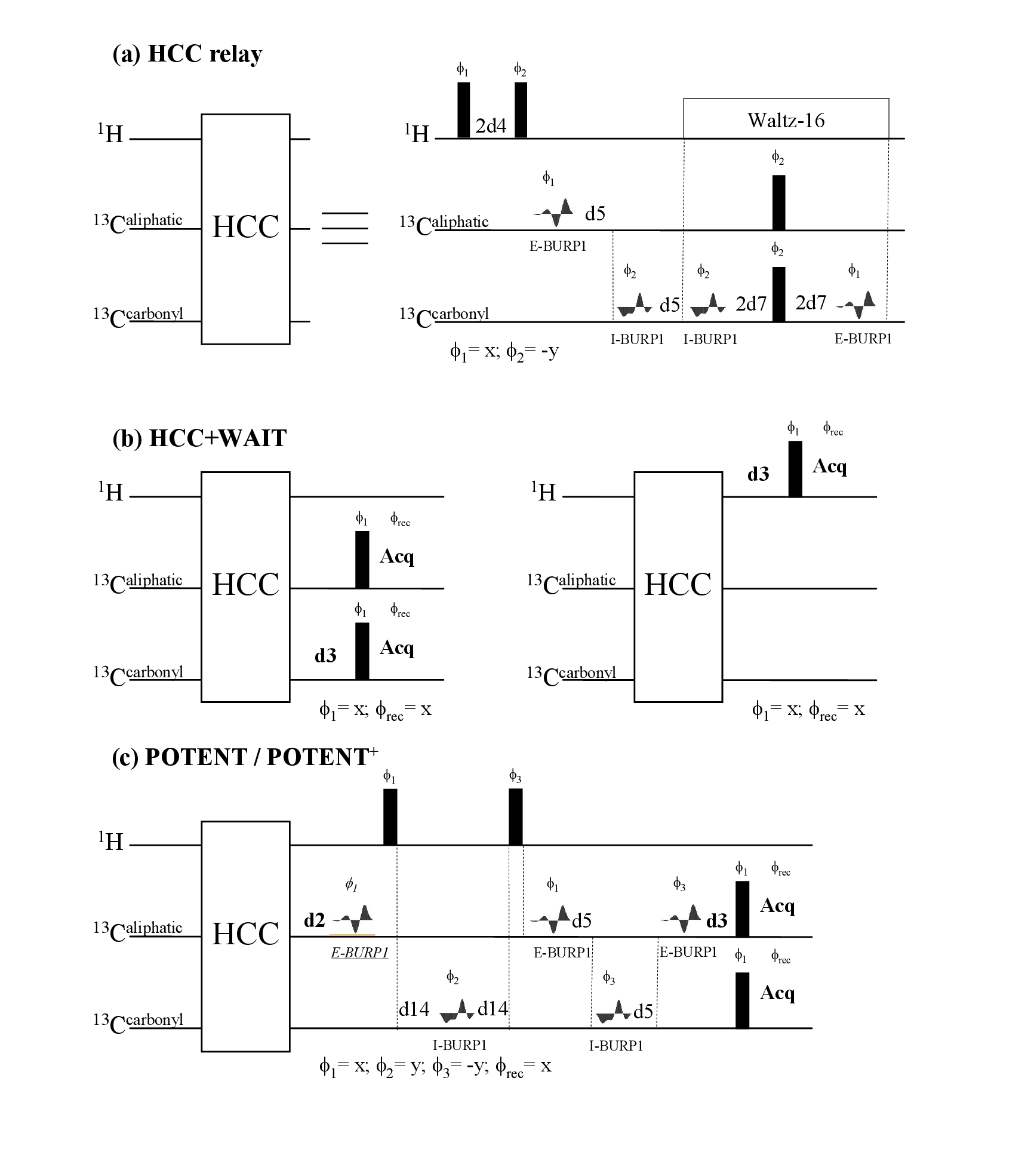}
\end{center}
\caption{
Pulse-sequences for heat-bath cooling of amino acids. Narrow black bars denote
$90^\circ$ non-selective pulses, while E-BURP1 and I-BURP1 represent
1~\msec\ spin-selective BURP pulses~\cite{GF91}. (a) The HCC relay was
 implemented by two refocused INEPT sequences. (b) A significant
delay ($d3 \sim 7\Tone(\Htwo)$) following the relay restored the polarization of
all protons. (c) Alternatively, both carbons were cooled by truncated POTENT
 sequences, where POTENT$^+$ includes an additional E-BURP1 $90^\circ$ pulse
 (highlighted) on \Ctwo\ after $d2$. Acquisition (after $90^\circ$ pulse) is
performed for both nuclei (b) or only for \iso{C}{13} (c). In figure (c) we
denote the pulse sequences as POTENT/POTENT$^+$ because for long $d3$ the proton
would repolarize (step 4 of POTENT).
}
\label{fig:pul}
\end{figure}

%
Detailed pulse sequences are illustrated in Fig.~\ref{fig:pul}. 
The HCC relay (Fig.~\ref{fig:pul}a) includes delays for the two
 INEPTs in step 1 ($d4$ and $d7$), and a variable refocusing period, $d5$,
set to be $\nicefrac{d4}{k}$ depending on the number, $k,$ of \Alp\
 protons~\cite{BE80}; For glutamate, $d5$=1.21~\msec\ ($d4$ = 1.71~\msec, $d7$ =
 4.681~\msec), while for glycine $d5$ = 0.39~\msec\ ($d4$ = 1.79~\msec, $d7$ = 4.742~\msec),
where $d5=\nicefrac{d4}{k}-\nicefrac{t}{2}~\msec,$ and the shortening (by
 $\nicefrac{1}{2}~\msec$) compensates for the refocusing I-BURP1 pulse on \Cone.    
The HCC+WAIT pulse sequence (Fig.~\ref{fig:pul}b) employed a long repolarization
delay, $d3 \sim 7\Tone(\Htwo).$
The delay $d7$ was optimized (around $1/4J_{CC}$) in $0.01~\Hz$
 intervals to maximize the integral of \Cone\ after HCC+WAIT.
 For POTENT (Fig.~\ref{fig:pul}c), the first
 repolarization delay, $d2 \gtrsim 2\Tone(\Htwo),$ allowed 
 significant cooling of \Ctwo\ by the following refocused INEPT.
 The final delay, $d3,$
 was of the order of \Ttwo(\Htwo) to eliminate small distortions, and not for
proton repolarization. Thus, the resulting POTENT is truncated.
For sample $E_M',$ in addition to truncated
 POTENT, we also
applied (truncated) POTENT$^+,$ where $d2$ was followed by a $90^\circ$ pulse on
 \Ctwo, resulting in uniform lines~\cite{BE80}.  
The delay $d14$ in PT \#2, step 3 of POTENT, was $d4-\nicefrac{t}{2}$ (again
shortened to compensate for the soft pulse I-BURP1).

%
 A single scan was acquired for the samples in Table~\ref{tab:t1}.
 In order to obtain the proton
 spectrum (when cooling the entire spin system), the HCC+WAIT sequence was
 repeated under the same conditions, 
while acquiring the proton spectrum (see Fig.~\ref{fig:pul}b). The NMR data were
apodized with
 small ($0.3\ \Hz$) exponential line broadening function prior to Fourier
 transformation.
The spectra were phase-corrected according to the equilibrium spectra, and
polarization enhancements were calculated from peak integrals.

\section{Results} 

\subsection{\Tone\ measurements}\label{sec:T1}
%
%
To affirm that \multiso{C}{13}{2}-labeled amino acids are suitable for
heat-bath cooling, we determined the longitudinal relaxation times
 (Table~\ref{tab:t1}) for glycine, sample $G$, and glutamate, sample $E$,
at room temperature ($24.0^{\circ}$C).
As expected, \Cone\ had much slower relaxation, resulting in \Tone\ ratios
of $\RR(\Cone, \Htwo) \sim 10,$ while $\RR(\Ctwo, \Htwo) \sim 1.5$ (see
 Table~\ref{tab:RR}).
 
%
For glycine in \DW\ (sample $G$), a long relaxation time was found for
 the carbonyl, $\Tone(\Cone)= 31.6 \pm 0.5 \sec,$ similar to the relaxation time
 recently reported for an aqueous solution of 1,2-\multiso{C}{13}{2}-glycine
 (16\% \DW), $\Tone(\Cone)^{aq}= 27.3 \pm 1.2 \sec$, obtained under similar
 conditions~\cite{GLE09}. For \Ctwo\ and the
 \Alp\ protons of glycine, we found much more rapid relaxation:
 $\Tone(\Ctwo)= 3.75 \pm 0.05 \sec,$ in good agreement with the aqueous
 solution, $\Tone(\Ctwo)^{aq} = 4.0 \pm 0.2 \sec$~\cite{GLE09} and
 $\Tone(\Htwo)= 2.72 \pm 0.02 \sec$. 

%
For glutamate (sample $E$), shorter relaxation times were found,
$\Tone(\Cone) = 13.03 \pm 0.04 \sec,$
 somewhat similar to an aqueous solution, $\Tone(\Cone)^{aq} = 10.2 \pm 0.8
 \sec$~\cite{vHLMG07}, 
 $\Tone(\Ctwo) = 1.96 \pm 0.02 \sec$ and $\Tone(\Htwo) = 1.29 \pm 0.02 \sec.$
 The relaxation times of the methylene protons were similar to $\Htwo$ (see
 Table~\ref{tab:t1}).

%
 Addition of \Mag\ to glutamate, sample $E_M,$ decreased all \Tone\
 relaxation times. Heating glutamate with \Mag\ to
 phys\-io\-logi\-cal temperature ($37.0^{\circ}$C), sample $E_M'$, increased all
 \Tone\ relaxation times by about $40\%$.
In both cases the resulting \Tone\ ratios were similar to sample $E,$ see
Table~\ref{tab:RR}.

\subsection{Cooling the spin systems of glycine and glutamate}
\begin{figure}[h]
\begin{center}
\includegraphics[scale=0.5]{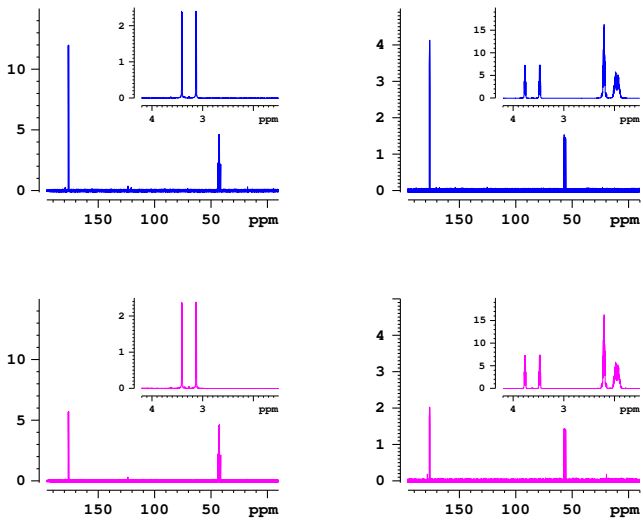}
\end{center}
\caption{
Cooling beyond Shannon's bound. \iso{C}{13} spectra (proton spectra in
 insets) of glycine (left, sample $G$) and glutamate (right, sample $E$) before
(bottom) and after (top) the HCC+WAIT experiment. \Cone\ was cooled by a factor
 of $\sim1.9,$ while the protons fully repolarized and \Ctwo\ regained its 
equilibrium polarization. 
}
\label{fig:HCCw}
\end{figure}

%
%
Our goal here was to reduce the total entropy of each amino acid spin system,
 thereby \emph{bypassing Shannon's bound} (for similar calculations see
 Ref.~\cite{POTENT}).
Both spin systems $G$ and $E$ were cooled at room temperature by a
single selective-reset, which consisted of an HCC relay that increased the
 signal of \Cone\ by a factor of about $3,$ followed by a
 significant delay, $d3 \sim 7\Tone(\Htwo).$ 
The pulse sequence HCC+WAIT (Fig.~\ref{fig:pul}b) includes two refocused INEPT
 sequences and employs proton decoupling during the carbon-carbon
 PT, similar to an earlier sequence~\cite{PGJ99}.
\Cone\ retained an enhancement factor of about $1.9$ ($1.88 \pm 0.01$ for
 glycine, $1.89 \pm 0.01$ for glutamate), such that the total entropy decreased
 by $\sim 2.6 \eps^2 / ln4$ (resulting from $2.6 \sim {1.9}^2 - 1^2$), where
 $\eps$ is the
 equilibrium \iso{C}{13} polarization. The protons repolarized back to
 equilibrium (within integration error $\lesssim1\%$) and \Ctwo\ returned to
 $1.00\eps \pm 0.01 \eps,$ see Figure~\ref{fig:HCCw}, suggesting that elementary
 heat-bath cooling is sufficient to go beyond the entropy bound. In terms of
effective spin temperatures, obtained by taking the quotient of the
 equilibrium temperature and the cooling factor (as done
in Ref~\cite{POTENT}), \Cone\ was cooled to $\sim156K$--$158K,$ while the other
spins were at the equilibrium temperature of $297.0\pm0.1K.$

%
For glutamate in the presence of \Mag\ (sample $E_M$), the HCC+WAIT
 sequence produced similar cooling as for sample $E$. \Cone\
 was cooled by a factor of about $1.9,$ while the other spins regained their 
equilibrium polarizations.

\subsection{Heat-bath cooling of \Cone\ and \Ctwo}
%
%
A potentially useful application of heat-bath cooling
is to enhance, simultaneously, two labeled carbons by applying the POTENT pulse
 sequence~\cite{POTENT} with no final delay; Fig.\-~\ref{fig:pul}c. For both
 amino acids (samples $G$ and $E$),
 \Cone\ and \Ctwo\ were cooled to a similar extent (about $2.4$-fold, see
Table~\ref{tab:POTENT}) when the first delay was $d2
 \sim 2$--$3 \Tone(\Htwo).$ The very short final delay ($d3 \lesssim 1
 \sec$) was used \emph{only} to eliminate small distortions. 
The spin temperatures after cooling were $\sim120K$--$130K.$

%
For glutamate with \Mag\ (sample $E_M$), application of steps 1--3 of POTENT
 resulted in
 cooling factors of about $2.5$ and $2.3$ for \Cone\ and \Ctwo, respectively,
 similar to sample $E,$ reflecting the
 similarity in \Tone\ ratios for both samples (see Table~\ref{tab:RR}).

\begin{figure}[h]
\begin{center}
\includegraphics[scale=0.48]{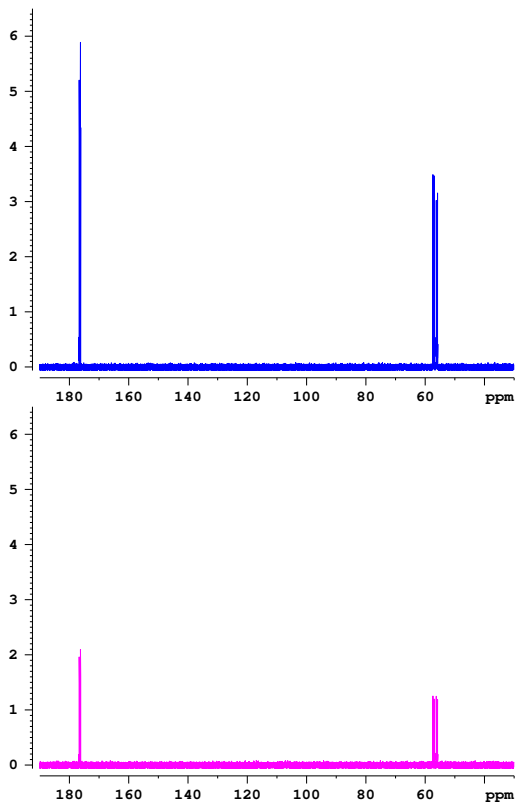}
\end{center}
\caption{
Simultaneous cooling of both backbone carbons of glutamate at physiological
temperature with Magnevist (sample $E_M'$) using POTENT$^+$, see
 Table~\ref{tab:POTENT} for details. Spectra are at equilibrium (bottom) and
 after cooling (top).
}
\label{fig:EGd310}
\end{figure}

%
Application of POTENT to glutamate at physiological temperature of
 $37.0^{\circ}$C (sample $E_M'$) cooled both carbons by a factor of about $2.5,$
quite similar to the cooling obtained at room temperature described above, see
bottom part of Table~\ref{tab:POTENT}.
Interestingly, improved cooling at physiological temperature is obtained by
POTENT$^+$ (Fig.~\ref{fig:pul}c), where an additional $90^\circ$ pulse was
 applied to \Ctwo, in order to provide uniform net PT (equal intensity for the
 lines of the \Ctwo\ multiplets~\cite{BE80}), see Fig.~\ref{fig:EGd310}.
 This sequence allowed a two-fold reduction in the final
delay ($d3$ of the order of \Ttwo(\Htwo)) by transforming erroneous
 density-matrix elements
 into non-observable coherences; consequently, for $d2=2.0$ $\Tone(\Htwo),$
 \Cone\ and \Ctwo\ were cooled by factors of about $2.60$ and $2.65,$
 respectively.

\section{Discussion} 

\subsection{Heat-bath cooling of amino acids for in vivo spectroscopy}
%
%
The simultaneous cooling of both labeled carbons at physiological conditions
 (sample $E_M'$) suggests that heat-bath cooling is suitable for 13C-MRS, which
 detects and quantifies \iso{C}{13}-labeled meta\-bo\-lites in vivo, in
 particular at specific locations within the brain~\cite{RC05}. For glutamate,
 it was recently shown that the INEPT pulse sequence can be adapted to in vivo
 13C-MRS and used for brain
spectroscopy with WALTZ decoupling~\cite{GAC+03, XS06}.
 Heat-bath cooling sequences that combine these elements may therefore be
 readily adapted to localized in vivo spectroscopy.

%
Simultaneous cooling of both \iso{C}{13} of 1,2-labeled amino acids by the
 POTENT pulse sequence may enhance the detection of such isotopomers in vivo.
 For glutamate, both \Ctwo-labeled and 1,2-labeled isotopomers are produced in
 cerebral meta\-bo\-lism of 1,2-\multiso{C}{13}{2}-glucose~\cite{RC05}. In this
 case, application of POTENT in vivo is expected to cool \Ctwo\ in both
 single and double-labeled isotopomers by means of the final PT
 (see Fig.~\ref{fig:pul}c).
Quantitative analysis may be facilitated by a suitable setting of $d2,$ such
 that both \iso{C}{13} of the double-labeled isotopomer are cooled to a 
similar extent, as shown in Table~\ref{tab:POTENT}.

%
Heat-bath cooling of the entire spin system as done here is a preliminary step
towards AC, which also demonstrates an important concept (in this case on
 important bio\-molecules):
On the one hand, reversible polarization compression can cool carbons of the two
 amino acids further, beyond just reaching the polarization of the sensitive
nucleus;
On the other hand, heat-bath cooling enables entropy reduction, which is not
attainable by polarization compression.
 
\subsection{Algorithmic cooling of amino acids}
%
%
Combining heat-bath cooling and compression provides AC that can, in theory,
 cool one or
 more spins far beyond the entropy bound~\cite{BMRVV02}. In reality, finite
 \Tone\ ratios limit the attainable cooling~\cite{EMW11, BEMW11}.
On the route to useful AC, one could employ various other amino acid
 isotopomers, such as 2,3,4-\multiso{C}{13}{3}-glutamate observed in
 vivo~\cite{DKH06} or 1,2-\multiso{C}{13}{2},\iso{N}{15}-N-acetyl
 aspartate (NAA). NAA, a major metabolite, is present in the brain at low
 concentrations and was detected in vivo by 13C-MRS~\cite{RLH+03, RC05,
 DKH06}. For this molecule, the amide proton constitutes a second
 reset spin and the labeled nitrogen (observed in vivo in the brain for
 glutamate~\cite{KR99}) serves as a computing spin. 
In both molecules, all computing spins may be significantly enhanced by
 parallel PT steps (a single reset step is required for NAA).
A 3-bit compression step could then cool one of the computing spins beyond
the proton polarization. 
For both molecules, further labeling would enable various practicable
 cooling algorithms~\cite{FLMR04, EFMW06, EMW11}. 
AC could ideally cool one of the 3 (6, 7) computation spins of fully-labeled 
glycine (glutamate, NAA), using a single reset spin, by a factor of 4 (32, 64)
 on top of the PT enhancement~\cite{SMW05, EFMW06}. 

%
Recently, in vivo 13C-MRS experiments with hyperpolarized biomolecules 
(e.g., \iso{C}{13}-pyruvate) have demonstrated the exciting potential of
 dissolution DNP and PHIP for \iso{C}{13} spectroscopy and imaging~\cite{GP06,
 KBNV08}. However, the time scale of metabolic processes within the brain
 imposes a severe limitation on the effectiveness of ex-vivo
 hyperpolarization, which is currently limited to few minutes at
 most~\cite{GP06, RBW+10}. The glutamine-glutamate cycle~\cite{RBHS03}, for
 example, between neurons and glia cells, is typically followed by 13C-MRS
 over the course of one to several hours~\cite{RLH+03, GAC+03, LSER03}.
AC of suitably labeled
 meta\-bo\-lites could be applied at any stage (with no additional equipment)
 and replenished in vivo, without harming the living tissue. For extended
 metabolic processes in the brain, a
 modest cooling (as could be obtained by AC)
of around $5$--$10$ fold is feasible in the near future. 
As a linear relationship between $1/\Tone$ and \Mag\ concentration was
shown for backbone carbons of both amino acids~\cite{GLE09, vHLMG07}, another
 interesting future avenue is to run AC with much higher concentration of \Mag\
 (e.g., $1\ \mM$).
 
%
In solid state NMR, AC was achieved for malonic a\-cid~\cite{BMR+05, RMBL08},
 using spin-diffusion (instead of thermalization) as a fast repolarization
mechanism for the protons, to enable up to four cycles of AC~\cite{RMBL08}.
 Very recently, several cycles of AC were achieved in solution for TCE, using
numerically optimized pulses~\cite{AEMW11}. In the far future, use of electrons
 as reset spins could enable application of AC to larger spin systems to obtain
 highly polarized spins~\cite{FLMR04, EMW11}.

\section{Conclusion}
%
%
Heat-bath cooling of the backbone carbons of glycine and glutamate may be
readily applied in vivo, and could also be applicable to other amino
 acids (and similar meta\-bolites). Going beyond heat-bath cooling, various
suitably labeled isotopomers of amino acids and similar metabolites may be
 cooled effectively by AC (potentially using \Mag).
Heat-bath cooling in the near future, and AC later on, may enhance spectroscopy
 of slow metabolic processes, particularly in the brain where hyperpolarization
 is not effective. These methods are safe, require no additional equipment, and
may be replenished.  

\section*{Acknowledgments}
We thank Dr Dorith Goldsher, Dr Yael Balasz and Yosi Atia for enlightening 
discussions, and we also thank D.G. for a sample of \Mag. This work is supported
in part by the Wolfson Foundation and the Israeli MOD Research and Technology
 Unit. The work of T.M. was also supported in part by FQRNT through INTRIQ, and
 by NSERC.

%
%


\twocolumn
\begin{table}[h]
\begin{minipage}{\textwidth}
\caption{%
\Tone\ values (in seconds) of \iso{C}{13}-labeled glycine and glutamate in \DW\
($\sim300\ \mM).$ Samples $E_M$ and $E_M'$ contain $0.05\ \mM\ \Mag.$
Errors are standard deviations of 3--10 single scans.
}~\\
\label{tab:t1}
\setcounter{mpfootnote}{\value{footnote}}
\renewcommand{\thempfootnote}{\alph{mpfootnote}}
\begin{tabular}{clclccccc}
\hline
sample &
Amino acid\footnote{Molar ratios of K$_3$PO$_4$ to amino acid were $0.5$ for
 glycine (pH 8) and $0.8$ for glutamate (pH 7.5).} &
T[K] & \Tone\ (\Cone) & \Tone\ (\Ctwo) &
\Tone\ (\Htwo) & \Tone\ (\Hthree) & \Tone\ (\Hfour)\\
\hline
$G$ &
glycine & $297$ & $31.6 \pm 0.5$ & $3.75 \pm 0.05$ & $2.72 \pm 0.02$\\
$E$ &
glutamate & $297$ & $13.03 \pm 0.04$ & $1.96 \pm 0.02$ & $1.29 \pm 0.02$ & $1.001 \pm 0.006$ & $1.281 \pm 0.003$\\
$E_M$ &
glutamate & $297$ & $10.2 \pm 0.1$\footnote{For sample $E_M,$ only two
measurements were taken for \Cone.} &
$1.84 \pm 0.02$ & $1.10 \pm 0.01$ & $0.920 \pm 0.003$ & $1.160 \pm 0.001$\\
$E_M'$ &
glutamate & $310$ & $14.36 \pm 0.06$ & $2.66 \pm 0.04$ & $1.50 \pm 0.01$ & $1.270 \pm 0.004$ & $1.606 \pm 0.004$
\end{tabular}
\setcounter{footnote}{\value{mpfootnote}}
\end{minipage}
\end{table}

\begin{table}[h]
\caption{%
\Tone\ ratios for the samples in Table~\ref{tab:t1}. 
}~\\
\begin{minipage}{\columnwidth}
\begin{tabular}{lcccc}
\hline
sample & \Mag[mM] & $\RR(\Cone,\Htwo)$ & $\RR(\Ctwo,\Htwo)$\\
\hline
$G$ & $0$ & $11.6 \pm 0.3$ & $1.38 \pm 0.03$\\
$E$ & $0$ & $10.1 \pm 0.2$ & $1.52 \pm 0.04$\\ 
$E_M$ & $0.05$ & $9.3 \pm 0.2$ & $1.67 \pm 0.03$\\
$E_M'$ & $0.05$ & $9.6 \pm 0.1$ & $1.77 \pm 0.04$
\end{tabular}
\end{minipage}
\label{tab:RR}
\end{table}

\begin{table}[here]
\caption{%
Simultaneous cooling of both backbone carbons by POTENT without (top two rows)
 and with added \Mag. In the last row, POTENT$^+$ was used (see
 Fig.~\ref{fig:pul}c and text).
 Results are averages and standard deviations of at least 3
single-scan measurements.
}~\\
\begin{minipage}{\columnwidth}
\setcounter{mpfootnote}{0}
\renewcommand{\thempfootnote}{\alph{mpfootnote}}
\begin{tabular}{clcccc}
\hline
sample &
$d2$/\Tone(\Htwo)\footnote{See Fig.~\ref{fig:pul}$c$.} &
$d3[\sec]\mathrm{^{\alph{mpfootnote}}}$ &
\Cone\ factor\footnote{The ratio between signals with and without cooling.} &
\Ctwo\ factor$\mathrm{^{\alph{mpfootnote}}}$\\
\hline
$G$ &
$2.2$ & $1.0$ & $2.32 \pm 0.06$ & $2.52 \pm 0.09$\\
$E$ &
$3.1$ & $1.0$ & $2.45 \pm 0.01$ & $2.29 \pm 0.02$\\
\hline
$E_M$ &
$1.8$--$2.7$ & $1.0$ & $2.45 \pm 0.08$ & $2.29 \pm 0.04$\\
$E_M'$ &
$2.7$ & $1.0$ & $2.51 \pm 0.04$ & $2.49 \pm 0.04$\\
$E_M'$ &
$2.0$
\footnote{With the delay used for POTENT ($d2=2.7\sec$), POTENT$^+$ cooled \Cone\ and
 \Ctwo\ by respective factors of about $2.5$ and $2.7.$} &
$0.5$ & $2.61 \pm 0.01$ & $2.65 \pm 0.01$\\
\end{tabular}
\setcounter{footnote}{\value{mpfootnote}}
\end{minipage}
\label{tab:POTENT}
\end{table}

\end{document}